# Control of polymorphism in coronene by the application of magnetic fields

**Authors:** Jason Potticary[1], Lui R. Terry[1], Christopher Bell[2], Andrew M. Collins[3], Claudio Fontanesi[4,5], Gabriele Kociok-Köhn[6], Simon Crampin[4], Enrico Da Como[4], Simon R. Hall[1*]

**Affiliations:**

[1] Complex Functional Materials Group, School of Chemistry, University of Bristol, Bristol, BS8 1TS, United Kingdom

[2] School of Physics, HH Wills Physics Laboratory, Tyndall Avenue, Bristol, BS8 1TL, United Kingdom

[3] Bristol Centre for Functional Nanomaterials, HH Wills Physics Laboratory, Tyndall Avenue, Bristol, BS8 1TL, United Kingdom

[4] Department of Physics, University of Bath, Claverton Down, Bath, BA2 7AY, United Kingdom

[5] Dipartimento di Ingegneria Enzo Ferrari, Universita' di Modena e Reggio Emilia, Via Vivarelli 10, 41125 Modena, Italy

[6] Department of Chemistry, University of Bath, Claverton Down, Bath, BA2 7AY, United Kingdom

*Correspondence to: simon.hall@bristol.ac.uk

**Summary**: Coronene, a polyaromatic hydrocarbon, has been crystallized for the first time in a different polymorph using a crystal growth method that utilizes magnetic fields to access a unit cell configuration that was hitherto unknown. Crystals grown in magnetic field of 1 T are larger, have a different appearance to those grown in zero field and retain their structure in ambient conditions. We identify the new form, β-coronene, as the most stable at low temperatures. As a result of the new supramolecular configuration we report significantly altered electronic, optical and mechanical properties.

Crystallization is a central theme in the preparation of materials and has a dramatic impact in medicine[1], biology[2], and materials science[3,4] as well as the food and manufacturing industries[5]. The continued development of novel drugs, proteins, and advanced materials strongly rely on our ability to self-assemble molecules in solids with the most suitable structure in order to exhibit desired functionalities[6]. In the field of functional molecular materials and pharmaceuticals it has been recently highlighted how polymorphism, i.e. the presence of different crystal structures in the same molecular system, can be an opportunity to discover novel phenomena[7] and tune properties[8]. For example, charge carrier transport in organic semiconductors can be increased by crystallizing molecules under pressure, resulting in shorter intermolecular distances favorable for transport[9].

The search for new polymorphs remains a scientific challenge that is at the core of crystal engineering[10]. One of the most successful approaches in this field is to perform small chemical modifications to a specific molecular skeleton in order to direct different intermolecular interactions. Unfortunately, this strategy necessarily modifies the molecular building blocks, influencing their electronic structure, which is often undesirable for applications where unmodified energy levels or chemical reactivity are crucial. In contrast, physical approaches rely on the use of pressure or different solvents (mixtures) in solution growth, or external fields leaving the chemical structure of molecules unaltered. Crystal growth in the presence of electric fields is a well-known process and is used extensively to prepare nonlinear optical materials through electric poling. It has also been used to selectively crystallise a known polymorph of glycine, through field-induced orientation of the highly polar glycine molecules[11]. Growth in the presence of an external magnetic field is much less explored however and consists mainly of the melt texturing of metals and alloys[12] and in cases where controlled convection of the crystallizing solution is required, for example in growth of high-quality protein crystals[13], chiral aggregates[14] or in the alignment of liquid crystal arrays[15]. While it is know that magnetic forces can have an effect on solidification, their use in changing the crystalline structure of molecular materials, i.e. to create new polymorphs in single crystals, has not previously been reported.

Here we show that by crystallizing the polyaromatic hydrocarbon coronene in the presence of a magnetic field, it can be made to form as a β-herringbone polymorph instead of the ubiquitous γ-herringbone form, with a change as large as 35º to the herringbone nearest neighbor angle. We critically analyze and discuss the role of the anisotropic magnetic susceptibility of the coronene molecule, and hence crystals, in driving the formation of this new polymorph. This anisropy is linked to the strongly delocalized π-electrons of coronene and its planar shape. The new β polymorph is stable and can preserve its structure in ambient conditions and zero magnetic field. Dispersion corrected density-functional-theory (DFT-D) calculations indicate that the new form is actually energetically favored at low temperature. Furthermore, we demonstrate how the new supramolecular structure generates remarkable changes of the electronic, optical and mechanical properties in the crystal.

**On the nature of polyaromatic hydrocarbon crystals**

Polyaromatic hydrocarbons (PAHs) are commonly researched molecules due to their rigid planar structure, high stabilities and characteristic optical and electronic behavior[16,17]. As molecular solids, PAHs crystallize in four basic structure types according to well-defined geometric and energetic considerations[18]. These are the herringbone (HB) structure, the gamma-herringbone (γ-) structure, the sandwich-herringbone (SHB) structure and the beta-herringbone (β-) structure (Extended Data Fig. 1). A comprehensive study of 32 PAHs has shown that the adoption of one of the four structure types depends ultimately on the relative strength of nonbonded C⋯C and C⋯H interactions[18]. By growing thin films of PAHs on substrates, epitaxial effects direct the formation of different polymorphs[19,20], although in single crystals at ambient pressure, polymorphism in PAHs is rare with perylene and pyrene being two notable cases exhibiting both HB and SHB polymorphs[21]. The ability to crystallize PAHs in different polymorphs under ambient pressure conditions would be a significant advance, as the physical properties of these molecules, e.g. semiconductivity, light absorption, carrier transport, and fluorescence are intimately related to the relative disposition of the polyaromatic molecules in the crystal structure.

Coronene is a PAH composed of six aromatic rings arranged in a planar discoidal geometry (Extended Data Fig. 2). It is a yellow crystalline solid at room temperature and occurs naturally as fibrous aggregates amongst sedimentary rocks (karpatite) in several parts of the world[22,23]. The high molecular symmetry ($D_{6h}$) and 24 electron π-system has made coronene an ideal model system for the study of graphene, due to it being large enough to display exotic electronic behavior, but not so large that contortion becomes a complicating factor[24]. Optical studies on coronene have suggested the presence of phase transitions at low temperature or high pressure, although their nature is currently unclear[25,26].

**Crystallization under magnetic field**

Centimeter-long crystals of coronene (typically around 0.75 cm) were grown from a supersaturated solution of the molecules in toluene cooled from 366 K (Fig. 1). Crystals of both polymorphs typically grew from solution when the temperature was between 328 K and 298 K over a period of 12 hours. Single crystal X-Ray diffraction (XRD) of crystals grown in the absence of magnetic field showed that the structure is the conventional γ-polymorph; $a = 10.02$, $b = 4.67$, $c = 15.60$, $\beta = 106.7°$, $Z = 2$, space group $P2_1/n$, with an inter-planar distance ($d_\pi$) of 3.43 Å, which is consistent with parallel π-stacking[18,27] (Fig. 2, right). In the presence of an external magnetic field of 1 Tesla however, significantly longer (typically around 2.5 cm) coronene crystals were grown from the same supersaturated solution and exhibit a different color to normal γ-coronene crystals (Fig. 1).

Single crystal XRD indicates that the structure of these crystals is consistent with a new β-coronene structure; $a$ = 10.39, $b$ = 3.84, $c$ = 17.23 Å, β = 96.24°, Z = 2, space group P2$_1$/n, and an interplanar distance ($d_\pi$) of 3.48 Å (Fig. 2, left). By comparing the two crystal forms in Figure 2, it can be seen that the short axis, $b$ is substantially decreased in length for the β-coronene crystallized in the magnetic field compared to the γ-coronene. This new polymorph also has a significantly smaller nearest neighbor herringbone angle of 49.71° compared to 95.86° in γ-coronene. The identification of the new crystal polymorph as a β-structure can be made by reference to a plot of the interplanar angle of nearest neighbor molecules versus the unit cell short axis[18] (Fig. 3). From this, it can be seen that the new polymorph sits squarely with other PAHs identified as having the β-herringbone structure. Growth performed at weaker field strengths of 0.2, 0.5 and 0.8 T resulted in crystals of the γ–polymorph (Extended Data Fig. 3), suggesting that 1 T is close to a threshold for energetic selection between the two forms. At 1 T, the new β-coronene polymorph grows exclusively, is reproducible (> 10 times to-date) and also occurs in the non-aromatic solvent hexane.

The two polymorphs reported here are textbook examples of how the physical properties of molecular solids are not only determined by their molecular electronic structure, but also by their molecular packing[28]. Fig. 1 shows optical images of the two different polymorphs taken under a UV lamp (λ = 365 nm). The remarkably different colors result from different light absorption characteristics. We measured the absorption spectrum of two single crystals of the two polymorphs as shown in Fig. 4A. This was obtained by shining unpolarized light perpendicular to the $a$-$b$ plane of the crystals, i.e. light propagation parallel to [001]. The γ-coronene single crystal is characterized by a first absorption resonance at 468 nm assigned to the free exciton in coronene, in agreement with previous studies[26]. The β-coronene spectrum is by stark contrast almost featureless, with an absorption onset at 780 nm and a maximum at ~500 nm. This is a remarkable change in the optical properties of this material. It is clear that β-coronene is capable of exhibiting greater light absorption over a much wider range of wavelengths than γ-coronene, leading to potential for use in organic photodiodes or solar cells. In order to explain the spectrum of the new β-coronene phase we first examined the bandgap using DFT+D calculations which correctly describe the existence of the new stable polymorph. These indicate that no appreciable difference in the indirect bandgap (Fig. 4 B-D) exists between the two polymorphs, suggesting that the large shift in the light absorption onset is instead related to a change in the fundamental photoexcitations in the two structures[29,30]. In β-coronene it is difficult to identify sharp resonances from bound Frenkel excitons as observed for the γ-polymorph. Instead, the structureless absorption band of the β–polymorph could originate from charge transfer (CT) excitons that are favored by the larger overlap between the molecular π orbitals. Moreover, the transition dipole moment of a CT exciton is likely to be oriented parallel to the *a-b* plane, which would ensure optimal coupling with the electromagnetic radiation in our experiment.

**Ring currents enable polymorphism**

Next we discuss the formation of β-coronene and the role of the magnetic field on the growth of the crystal. Firstly, based on inductively-coupled plasma atomic absorption spectroscopy, we note that the level of magnetically active impurities such as cobalt, iron and nickel are at the parts per billion level (Co = 0.138 ± 0.98 ppb; Fe = 33.81 ± 0.56 ppb; Ni = 13.46 ± 0.65 ppb), and so are unlikely to play any role in polymorph selection. Instead we note that an applied magnetic field induces two electronic ring currents in the aromatic electrons of coronene, with the inner ring giving a paratropic, and the outer rim a stronger diatropic circulation of electrons[31]. Consequentially a highly anisotropic magnetic susceptibility $\chi$ arises, with $\Delta\chi = \chi_\parallel - \chi_\perp = -8.1 \times 10^{-33}$ m³/molecule[32], where $\parallel$ and $\perp$ refer to directions parallel and perpendicular to the molecular axis $\hat{n}$. We believe that this strong magnetic anisotropy produced by the induced ring currents is the key to the effective coupling of the coronene molecules to an applied magnetic field of 1 T. In the case of other PAHs such as pyrene, which do not have such a strong induced ring current (as evidenced by a deshielded proton chemical shift of 8.0 ppm determined by NMR) and concomitant magnetic anisotropy as coronene (large deshielding of the proton signal at 8.9 ppm (viz. 7.0 ppm for benzene)), our attempts to control polymorphism at 1 T were unsuccessful. With higher applied fields, it is likely that polymorphism in pyrene and other aromatic systems can be realised.

In coronene, the induced electronic ring currents are expected to produce negligible changes to the intramolecular bonds for the field strengths used in our experiments, but in solution the applied field will tend to align the molecules so as to minimize their magnetic potential energy. The energy change associated with bringing a coronene molecule into a field $\vec{B}_0$, $\Delta E = -\vec{B}_0 \cdot \chi \cdot \vec{B}_0/2\mu_0$ favours molecular orientation with $\hat{n}$ perpendicular to $\vec{B}_0$, but only slightly; $\Delta E_\parallel - \Delta E_\perp = -B_0^2 \Delta\chi/2\mu_0 = 3.2 \times 10^{-27}$ J in a 1T field, leading to an expected degree of alignment $S$, with $\vartheta$ the angle between $\hat{n}$ and $\vec{B}_0$, of [33]

$$S = \langle \frac{3}{2}\cos^2\vartheta - \frac{1}{2}\rangle \approx \frac{\Delta\chi B_0^2}{15 k_B T \mu_0} = -0.9 \times 10^{-7}.$$

$S$ ranges from 1 for molecules fully aligned with the field, to $-1/2$ for perpendicular alignment, with 0 indicating random orientation. Thus we can exclude a preferential alignment of coronene molecules in the growth solution at 350 K in a 1 T field. The magnitude of dipolar interactions between induced molecular moments $\vec{m} = \chi \cdot \vec{B}_0/\mu_0$ can be estimated from $U \sim \mu_0 m^2/4\pi r^3$ and is $\sim 3 \times 10^{-32}$ J for coronene-coronene separations of 3.5 Å, completely negligible compared to thermal energy at growth temperatures, $k_B T \sim 5 \times 10^{-21}$ J.

Excluding orientational effects on the isolated molecules and magnetic dipole coupling leaves cooperative phenomena arising from aggregates of molecules. Because of their distinct molecular arrangements, the energy change in the presence of the magnetic field has a differing orientational dependence for the two polymorphs, whilst scaling with the number of molecules $N$ in crystalline aggregates. Quantifying the probability that the field will orient crystals of different sizes through the order parameter along three orthogonal directions (Extended Data Fig. 4), we find that for crystals containing small number of molecules the energy due to immersion in the field is so small that there is little ordering ($S \approx 0$ in all directions), and no distinction between γ– or β-coronene crystals arises. No distinction exists for very large crystals ($N \gtrsim 10^8$) either, because in this case the tendency to orient is very large for both polymorphs. Between these limits a regime exists when the energy scale associated with the orientation in the field ($\sim NB_0^2 \Delta\chi/2\mu_0$) is comparable to $k_B T$, and a physically significant distinction between γ–coronene and β-coronene crystalline structures arises. We estimate that this occurs around $N \sim 10^6$ in a 1 T field at 350 K. This difference appears (even though both polymorphs can orientate themselves so that their magnetic-field induced energy has the same minimum energy state) due to differences in the energy cost for misalignment away from this orientation, with the β-polymorph having higher energy costs (Fig. 5). Qualitatively this is clear by considering the more planar arrangement of the molecules in the *a-c* plane of β-coronene compared to the γ-polymorph (Extended Data Fig. 5). We speculate that kinetically this may favor the reorientation of β-clusters over the γ-, which may both be present in solution, enhancing the growth rate by cluster combination of the former above 1 T.

The elongated shape of both crystal forms in the *b* direction suggests that the rate of growth is largest along *b*. On the assumption that growth occurs from nucleation centers having an internal structure very similar to the formed γ– and β-crystals[34], the β– nucleation centers of a certain size have on average a more oriented *b*-axis than γ ones. This is consistent with the sphere plots in Fig. 5, where the lowest energy perturbation of the orientation for the β–phase occurs by leaving the *b* axis unaltered, i.e. equatorially in the energy sphere.

It is important to note that the appearance of polymorphs at later stages of growth, when there are many thousands of molecules ordered, is consistent with the Ostwald's step rule[35-37]. At such a stage the magnetic field may play a more significant role in altering the thermodynamic landscape. Applied magnetic fields have previously been found to affect the nucleation and growth rate of inorganic systems, proteins and paracetamol[38-40]. Enhanced growth rates induced by the magnetic field are especially notable; since there is strong evidence that the growth rate and not the nucleation rate can be the key kinetic factor in determining which polymorph dominates during crystallization[41,42].

**Polymorph stability**

The β-polymorph structure is notable for the large change in the herringbone angle. We can gain insights into the relative stability of the β-coronene structure when compared to the γ-polymorph, by consideration of the changes in the CH⋯π hydrogen bonding motif. It is known that the stronger the hydrogen bond, the stronger the trend for linearity[43,44] depending on the strength of the proton donor. From XRD, we can see that the CH⋯π angle in γ-coronene of 95.86° suggests a strong, almost linear CH⋯π hydrogen bond, whilst in β-coronene, this angle becomes 49.71°. The change in angle would suggest a weakening of the hydrogen bonding in β-coronene, which is concomitant with the change in estimated CH⋯π hydrogen bond length[45] that increases from 2.5 Å in the γ-polymorph, to 3.0 Å in the β-polymorph. This weakening of the CH⋯π hydrogen bonding should therefore manifest itself as a diminution of the physicomechanical properties of β-coronene. Indeed, we have determined the melting point ($T_M$) and elastic modulus ($E$ – the modal value measured on the ($\bar{1}011$ and $10\bar{1}\bar{1}$ crystal faces) of both polymorphs and find that in γ-coronene, $T_M$ = 436.43 °C ± 0.01 °C and $E$ = 227 GPa, whereas for β-coronene $T_M$ = 434.94 °C ± 0.01 °C and $E$ = 92 GPa. (Extended Data Fig. 6 and 7). These data suggest weaker inter-molecular forces in β-coronene at room temperature.

Finally, we undertook a series of DFT-D calculations, which suggest β-coronene is the more stable of the two polymorphs at 0 K, by 3.8 kJ/mol based upon lattice energy differences (Extended Data Fig. 8). Confirmation that the β-polymorph is the more thermodynamically stable comes from powder XRD of polycrystalline γ-coronene recorded through a temperature cycle from room temperature to 12 K. On cooling through a temperature of 150 K, three new peaks emerge which cannot be indexed to the γ–polymorph (JCPDS card number 12-1611) (Extended Data Fig. 9). These emergent peaks correspond to the (002), (101) and (112) ($2\theta$ = 10.55°, 10.67° and 27.72° respectively) reflections of the new polymorph, β-coronene. Thus β- and γ-coronene are enantiotropic polymorphs, with a critical temperature between 100 K and 150 K (Fig. 6).

In summary, through judicious application of a magnetic field, we have demonstrated for the first time that a well-studied polyaromatic hydrocarbon can be grown in a new polymorph and confirmed that a range of physical properties have been significantly altered as a result. Furthermore, simulation shows that this new polymorph is in fact a more thermodynamically stable form of coronene at low temperatures. The relatively low field strengths (1 T) used in our study holds promise for the discovery of new, unobserved polymorphs at higher fields in other aromatic crystal systems. This immediately suggests potential impact in areas as diverse as solid-state lasers, field-effect transistors and most tantalizingly, pharmaceuticals.

**Acknowledgments**

S.R.H. and J.P. acknowledge the Engineering and Physical Sciences Research Council (EPSRC), UK (grant EP/G036780/1), and the Bristol Centre for Functional Nanomaterials for project funding. In addition they would like to thank Stasja Stanišić for discussions on low-temperature X-Ray crystallography and Spartacus Mills for estimations of magnitude. E.D.C. and S.R.H. are grateful to the GW4 for funding through an accelerator grant and LRT is supported in part by a grant from the USAF European Office of Aerospace Research and Development (FA8655-12-1-2078). All authors would like to thank Prof. John Evans and Liana Vella-Zarb of the University of Durham for low-temperature powder XRD, and Dr Chung Choi and Dr George Whittell of the University of Bristol for ICP-AES and DSC respectively.

**Contributions**

S.R.H. initiated and supervised the project. J.P. performed the crystallisation experiments and characterised the samples optically and with C.B. undertook magnetic characterisations. L.R.T. performed DSC experiments, E.D.C. and C. F. absorbance spectra experiments, G.K.K. single crystal X-ray characterisation. S.C. undertook DSC calculations and with E.D.C, determination of band structures. A.M.C. performed elastic moduli experiments. All authors contributed to the discussion of the results, analysis of the structures and to manuscript preparation.

**Author Information** Original optical images and raw data from XRD, AFM and DFT-D calculations pertaining to all materials in this manuscript have been placed in the University of Bristol Research Data Repository ([www.data.bris.ac.uk/data/](www.data.bris.ac.uk/data/)). The crystallographic cif file for β-coronene has been lodged with the Cambridge Crystallographic Data Centre under the deposition number CCDC 1409823. The authors declare no competing financial interests. Correspondence and requests should be addressed to S.R.H. (simon.hall@bristol.ac.uk).

**Figures**

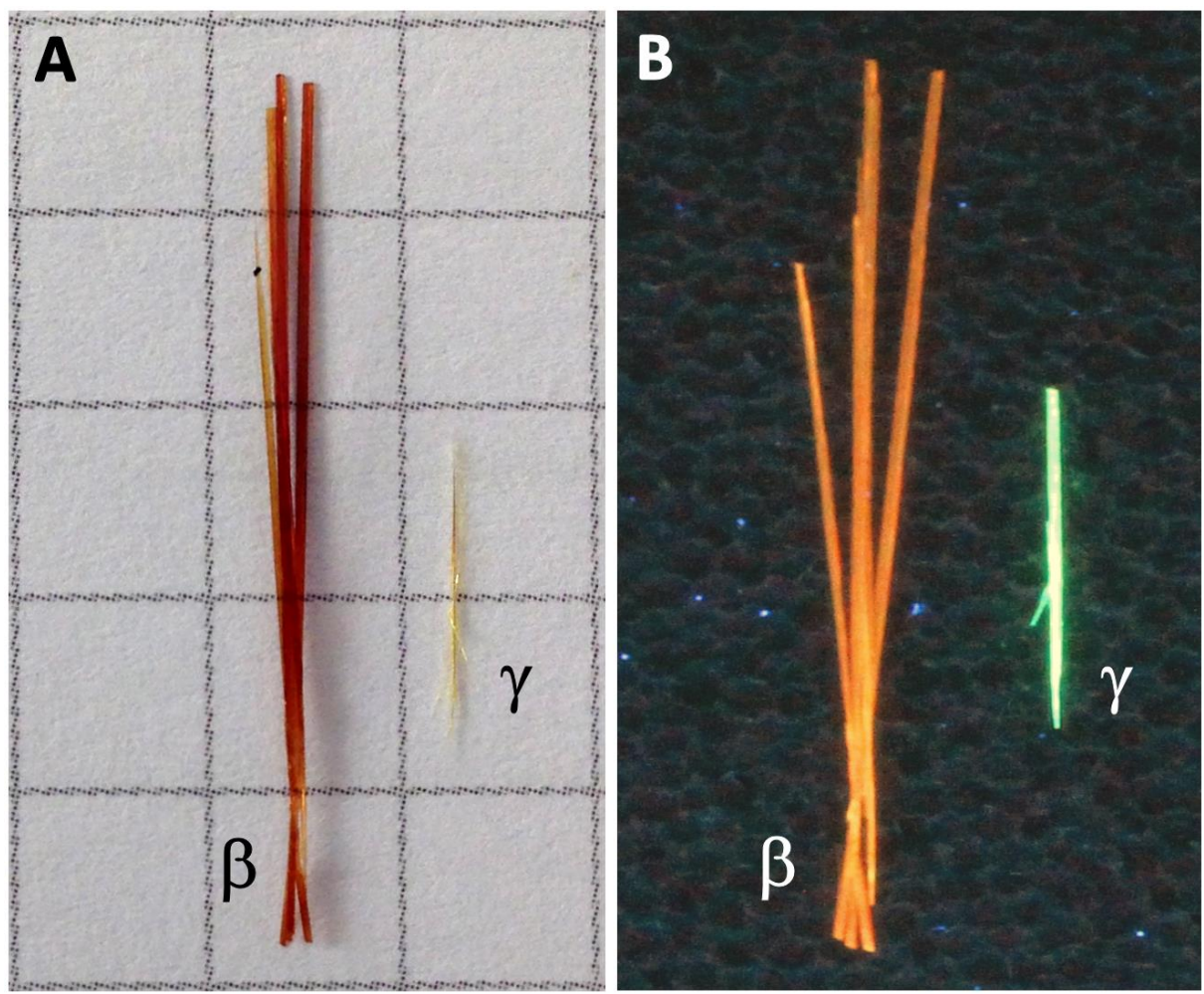

**Figure 1**. **Optical images of the β- and γ- polymorphs of coronene.** (A) in daylight and (B) under UV (λ = 365 nm) illumination. The squares on the grid in (A) are 0.5 cm × 0.5 cm.

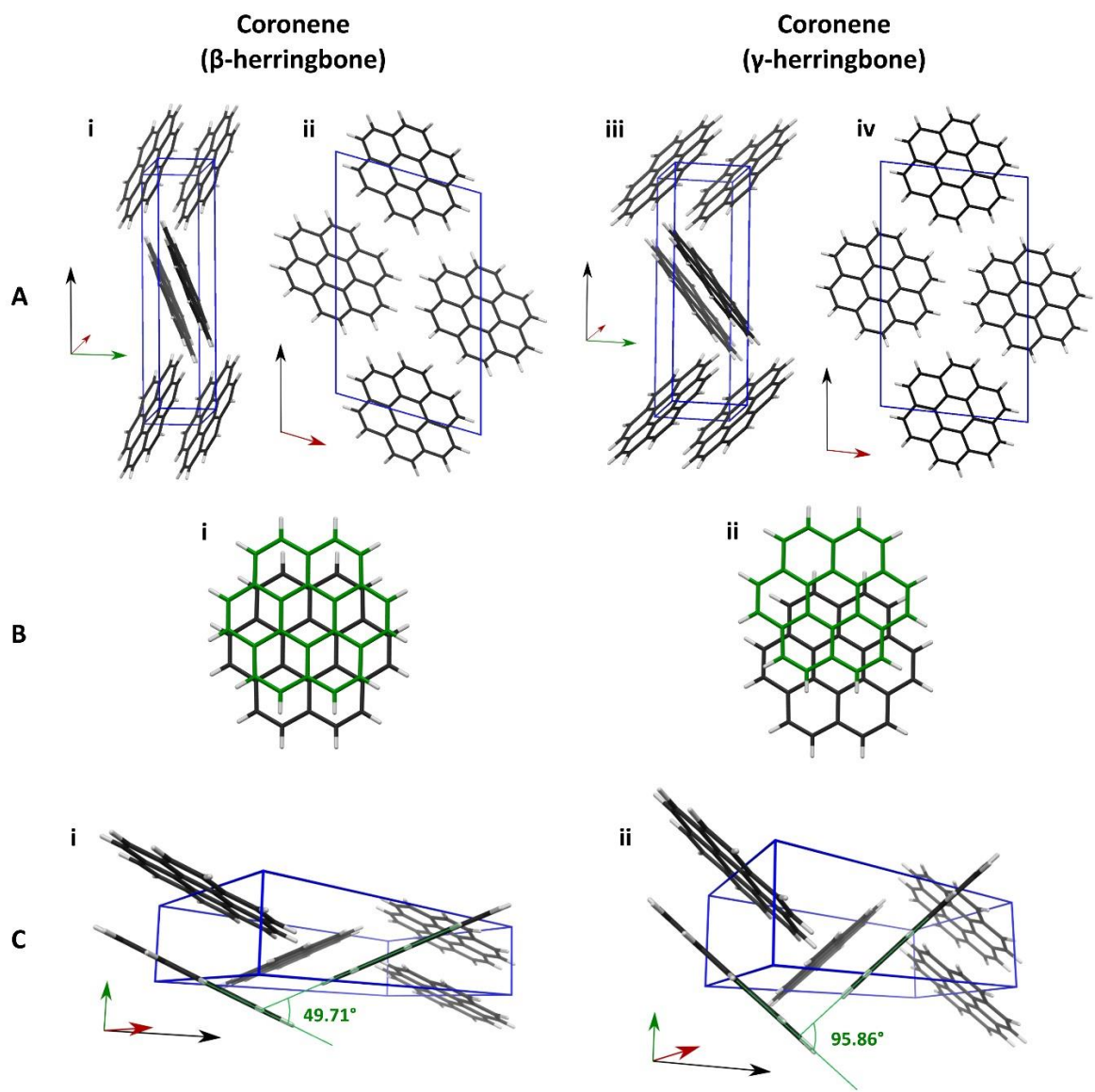

**Figure 2**. **Representation of the β- (left) and γ- (right) forms of coronene.** Differing perspectives of the unit cell **A** (blue box) viewed slightly offset from along the *a*-axis (**i** and **iii**) and along the *b*-axis (**ii** and **iv**). **B** shows the relative shift of the molecules along the stacks for β- (**i**) and γ- (**ii**). **C** shows an orientation of the unit cell clearly demonstrating the difference in nearest neighbor angle. Red green and black arrows indicate the direction of the *a*-, *b*- and *c*-axis respectively.

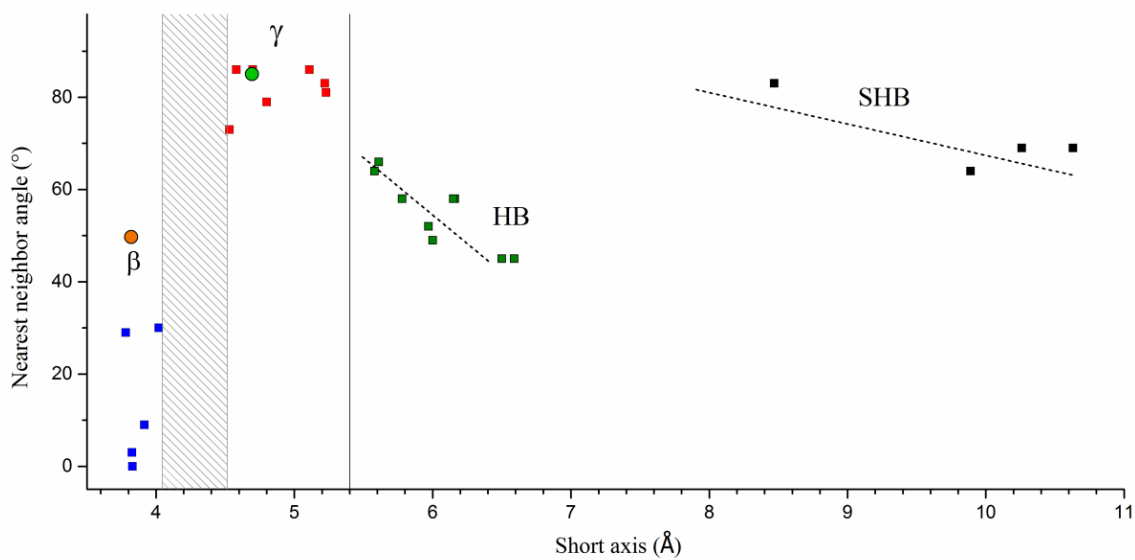

**Figure 3. Grouping of PAHs into structure types according to the crystallographic short axis and nearest-neighbor herringbone angle.** Plotted on the graph are PAHs that adopt the herringbone (HB) structure (green squares), the gamma-herringbone (γ-) structure (red squares), the sandwich-herringbone (SHB) structure (black squares) and the beta-herringbone (β-) structure (blue squares). The positions of both γ- and β-coronene polymorphs are indicated by circles. Adapted from ref. 18 in which the names of the crystals corresponding to all of the marked squares can be found.

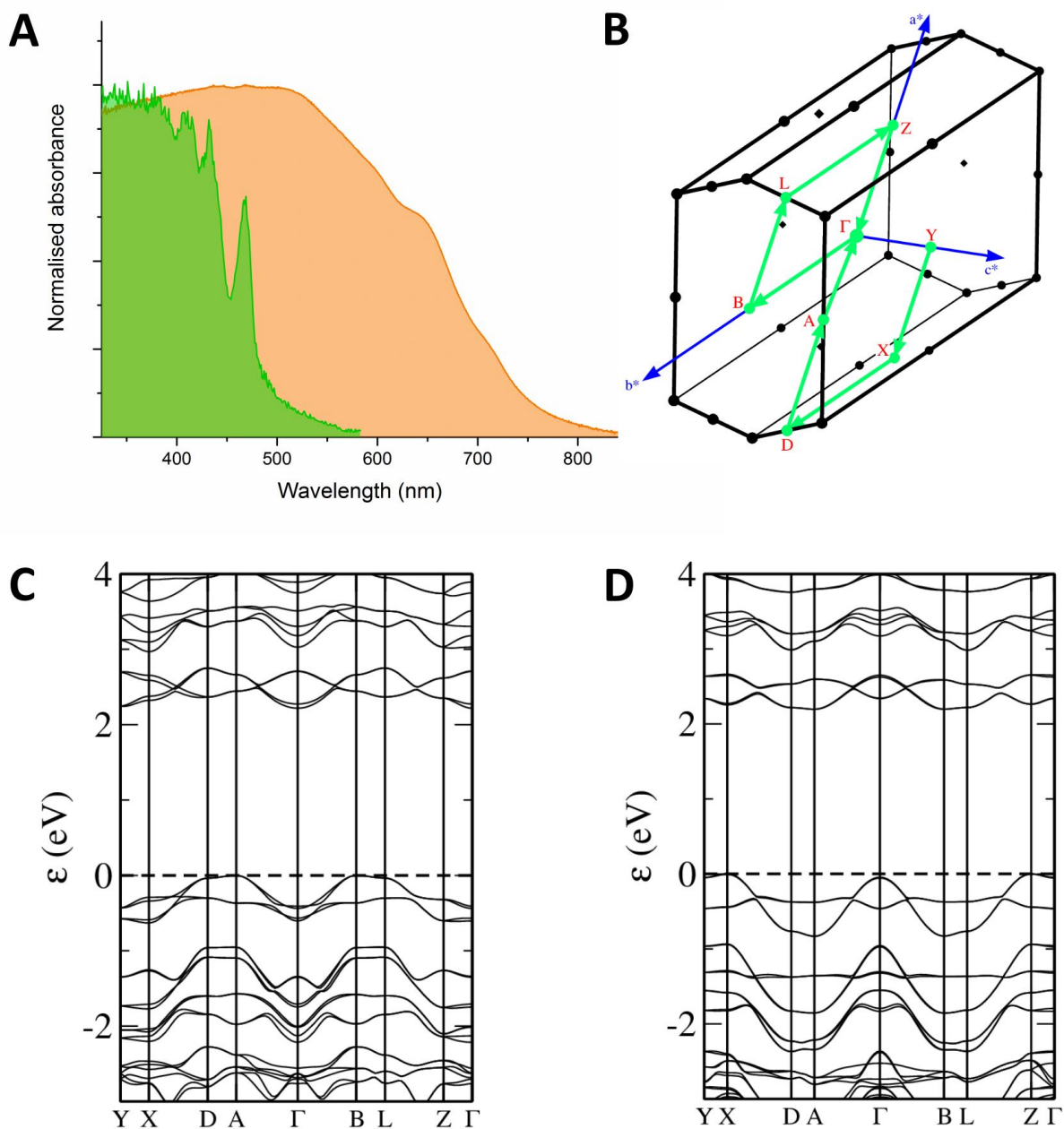

**Figure 4. Electronic structure of γ- and β-coronene.** (A) Absorption spectra of γ-coronene (green) and β-coronene (orange) single crystals. Unpolarized light was irradiated perpendicular to the *a-b* plane at room temperature; (B) Brillouin zone with reciprocal lattice vectors and high symmetry points; (C) and (D) band dispersion along high symmetry points in γ- and β-coronene, respectively.

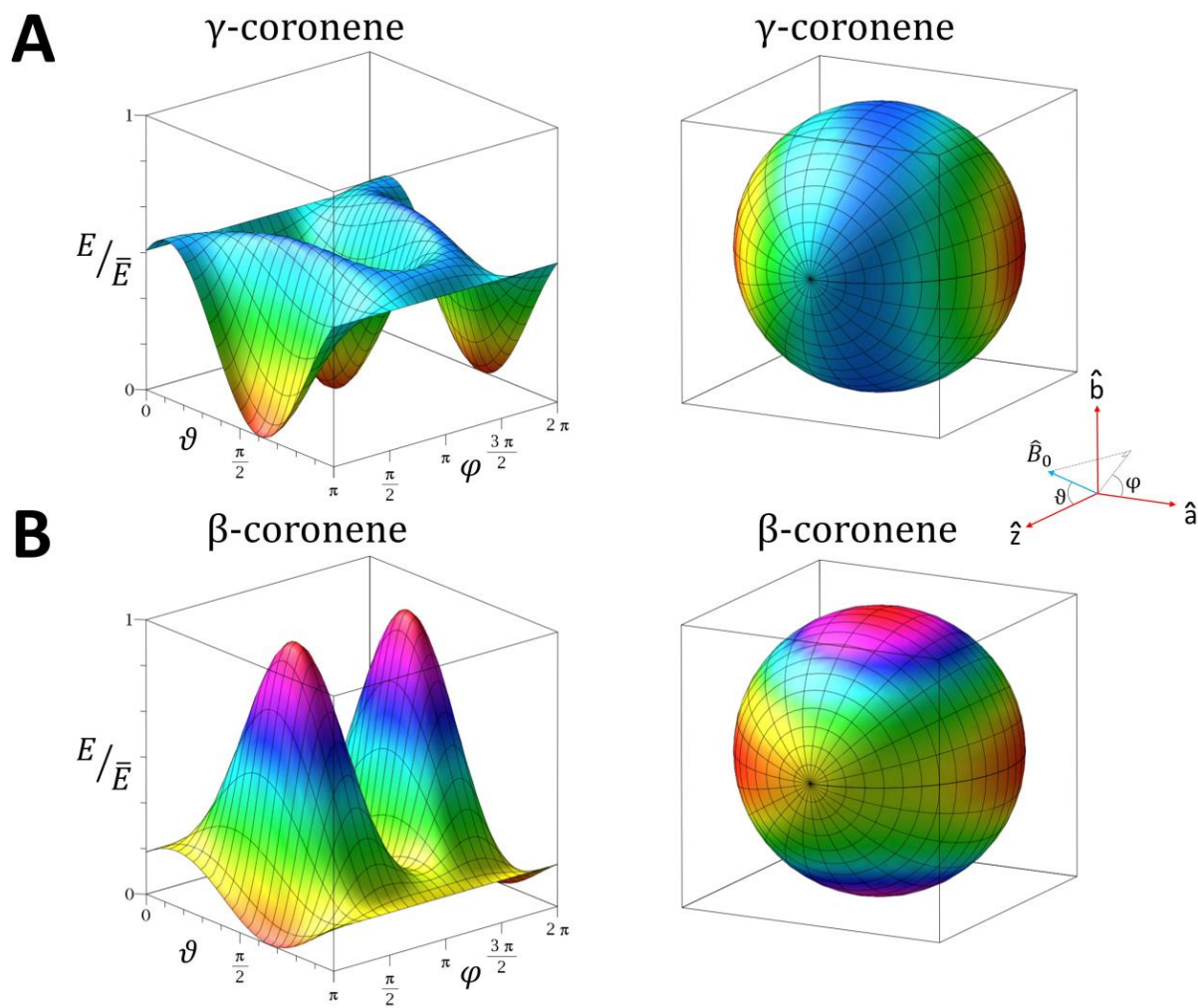

**Figure 5. Orientational dependence of the energy of γ- and β-coronene crystals in a magnetic field.** Calculated using the DFT-D geometries, the crystal is assumed fixed with *a* and *b* axes aligned along *x* and *y* and the magnetic field rotated, with ϑ and φ the usual spherical polar angles. The left panels show the energy as a fraction of $\bar{E}$, the energy of the most energetically unfavorable orientation of either crystal, which is $5.4\times10^{-27}$ J/T$^2$ per unit cell. The right panels show the same quantity as a color-coded sphere plot. The constant energy associated with the mean susceptibility is not included.

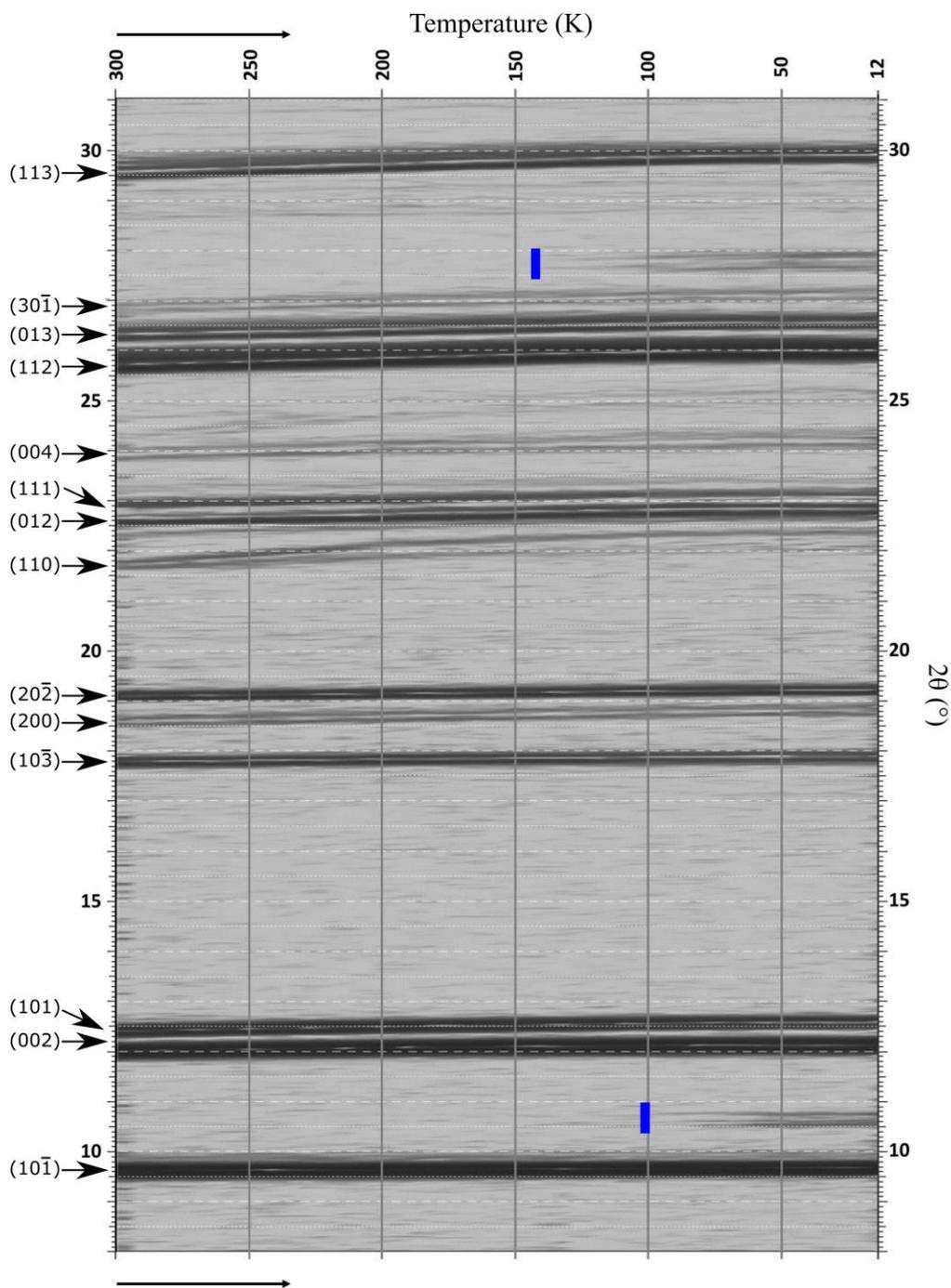

**Figure 6. 2D map of XRD peak shifts as a function of temperature in γ-coronene.** Temperature is decreased from 300 K to 12 K (left to right). Blue markers indicate the emergence of the new peaks due to β-coronene formation. The blue line near 10° 2θ encompasses two emergent peaks.

## Methods

**Coronene**. Coronene crystals (Sigma Aldrich, purity 97%) were recrystallized from toluene and purity assayed by NMR ($C^{13}$ and $H^+$). Trace metals were tested for using inductively coupled plasma – atomic emission spectroscopy (Agilent 710 ICP-AES) after digestion in 3% nitric acid (Sigma Aldrich, ≥99.999%, trace metal basis). Structural analysis was conducted by optical microscopy and single crystal x-ray diffraction (XRD). Analysis of the single crystal data of non-magnetic field grown crystals suggests favorable growth along the *b*-axis, which form the characteristic needle shaped crystals.

**Crystal growth under 1T of magnetic field.** A supersaturated solution of coronene (2.5 mg/ml) in toluene was prepared and stored in an oven at 93 °C. The solution was then passed through a 0.22 µm PTFE filter directly into a 5 mm quartz cuvette with a stopper. Once sealed, the cuvette was placed in the magnetic cavity (Extended Data Fig. 10) and the whole system maintained at 93 °C for 4 hours post cuvette insertion. The oven was then programmed to cool to 83 °C, 73 °C, 63 °C then finally 50 °C, with a 24 hour dwell at each temperature, before cooling to room temperature. Crystals of β-coronene are the sole polymorph to grow under the magnetic field and to-date the experiment has been repeated > 10 times.

**Single crystal X-ray crystallography.** Intensity data for all coronene structures were collected at different temperatures on an Agilent SuperNova-E Dual diffractometer equipped with an Oxford Cryosystem, using CuKα radiation (λ= 1.5418 Å). Data were processed using the CrysAlisPro software (CrysAlisPro, Agilent Technologies, Version 1.171.37.35 (release 13-08-2014 CrysAlis171 .NET) (compiled Aug 13 2014). For all structures a symmetry-related (multi-scan) absorption correction was applied. Crystal parameters are provided in (Extended Data Fig. 11) Structure solution, followed by full-matrix least squares refinement was performed using the WINGX-v2014.1 suite of programs throughout.

**Physical characterization.** Optical absorption spectra were recorded with a UV-VIS-NIR spectrometer Agilent Cary 5000 measuring in transmission configuration. The spectrometer can measure absorbance up an optical density of 10. Single crystals of β- or γ-coronene were suspended in the sample beam path. Unpolarised light was shone perpendicular to the *a-b* plane of the unit cells. High quality crystals with flat surfaces were carefully selected to avoid light scattering effects. The beam size was narrowed to half the lateral size of crystals. All absorption spectra were recorded at room temperature in air. Elastic moduli were determined using an Oxford Instruments MFP-3D Infinity AFM operating in AMFM viscoelastic imaging mode, using an AC160TS-R3 silicon tip (9 nm ± 2 nm radius). Freshly cleaved mica was used as a calibration standard (measured at 178 GPa). Melting point determination was done using a TA-Instruments Q100 DSC with temperature ramp of 10 °C min$^{-1}$ between 35 – 460 °C with 3.0 mg of coronene hermetically sealed in an aluminium pan.

**Computational calculations**. Computational calculations have been performed with density functional codes CASTEP[46] and VASP[47], using the Perdew-Burke-Enzerhof (PBE) exchange correlation functional with semi-empirical dispersion corrections (DFT-D) to account for van der Waals interactions. CASTEP calculations (version 7.03) used in-built ultra-soft pseudopotentials for C and H atoms, a plane wave cut-off of 600 eV, and Monkhorst-Pack[48] k-point samplings of 3 x

4 x 2. Cell parameters and atomic coordinates were fully relaxed using the BFGS method, halting when residual forces fell below 1 meV Å$^{-1}$. Changing the cut-off energy from 450 eV to 600 eV caused structural parameters to change by <0.001 nm or <0.01°. Increasing k-point sampling to 6 x 8 x 4 changed energies by < 1 meV. Robustness of results to choice of semi-empirical dispersion correction was assessed through the use[49] of using the Grimme scheme[50] with both default vdW radii ($R_H$ = 1.001, $R_C$ = 1.452) and experimental ($R_H$ = 1.090, $R_C$ = 1.750), as previously employed by Fedorov *et al.*[51] for γ-coronene. VASP calculations (version 5.3.3) use PAW potentials[52] with 500 eV energy cut-off, 3 x 4 x 2 k-point sampling and the Tkatchenko-Scheffler[53,54] dispersion correction. Geometry optimization was performed starting from experimental cell parameters and atom coordinates, both with and without symmetry constraint. Stability of optimized geometries was verified by re-optimizing after randomly displacing atoms by 0.05 Å in *x*, *y* and *z*.

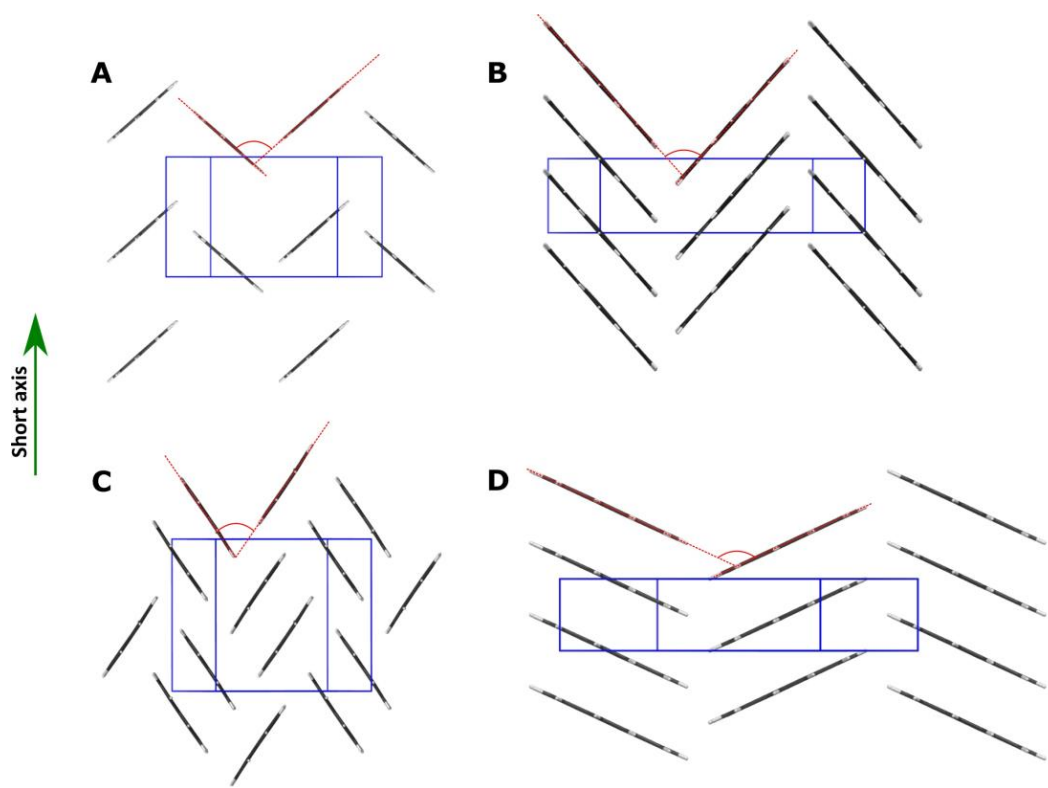

**Extended Data Figure 1. The four classes of crystal packing in polyaromatic hydrocarbons.** (A) is the herringbone (HB) structure, (B) the gamma-herringbone (γ-) structure, (C) the sandwich-herringbone (SHB) structure and (D) the beta-herringbone (β-) structure.

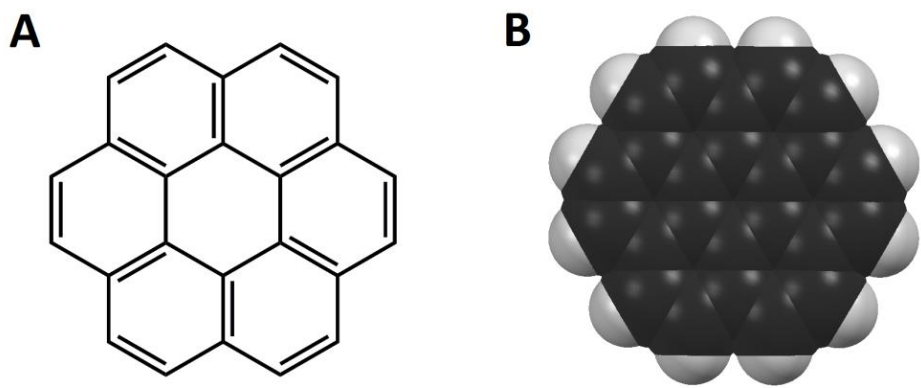

**Extended Data Figure 2. The molecular structure of coronene** shown as (A) skeletal and (B) space-filling models.

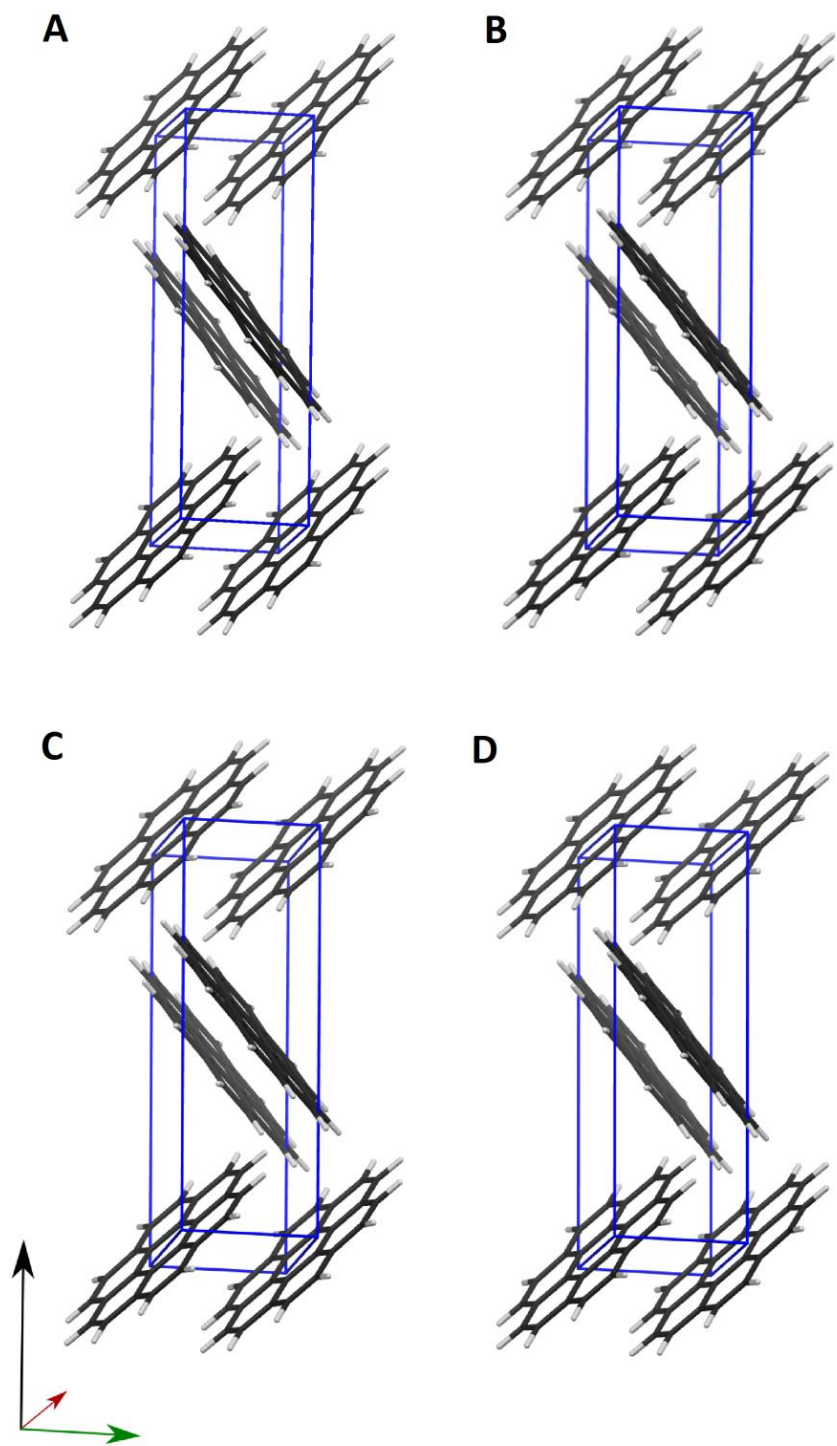

**Extended Data Figure 3. X-Ray diffraction crystal structures of coronene grown under a range of applied fields.** In (A) the applied field was 0 T; in (B) 0.2 T; in (C) 0.5 T and in (D) 0.8 T. Red green and black arrows indicate the direction of the *a*-, *b*- and *c*-axis respectively.
24

$$S = \langle f(\vartheta, \varphi) \rangle = \frac{\iint f(\vartheta,\varphi) e^{-E(\vartheta,\varphi)/kT} d\Omega}{\iint e^{-E(\vartheta,\varphi)/kT} d\Omega}$$

Szz=<3/2 cos²ϑ-1/2>

| N | γ | β |
|---|---|---|
| $10^3$ | 8.9×$10^{-5}$ | 8.9×$10^{-5}$ |
| $10^4$ | 8.9×$10^{-4}$ | 8.9×$10^{-4}$ |
| $10^5$ | 9.0×$10^{-3}$ | 8.9×$10^{-3}$ |
| $10^6$ | 0.094 | 0.087 |
| $10^7$ | 0.746 | 0.552 |
| $10^8$ | 0.977 | 0.959 |

Sxx=<3/2 sin²ϑcos²φ-1/2>

| N | γ | β |
|---|---|---|
| $10^3$ | -3.9×$10^{-5}$ | +4.4×$10^{-5}$ |
| $10^4$ | -3.9×$10^{-4}$ | +4.4×$10^{-4}$ |
| $10^5$ | -3.9×$10^{-3}$ | +4.3×$10^{-3}$ |
| $10^6$ | -0.042 | +0.033 |
| $10^7$ | -0.366 | -0.124 |
| $10^8$ | -0.488 | -0.466 |

Syy=<3/2 sin²ϑsin²φ-1/2>

| N | γ | β |
|---|---|---|
| $10^3$ | -5.07×$10^{-5}$ | -13.3×$10^{-5}$ |
| $10^4$ | -5.08×$10^{-4}$ | -13.3×$10^{-4}$ |
| $10^5$ | -5.10×$10^{-3}$ | -13.2×$10^{-3}$ |
| $10^6$ | -0.052 | -0.120 |
| $10^7$ | -0.379 | -0.429 |
| $10^8$ | -0.489 | -0.493 |

**Extended Data Figure 4**. Probability that crystals of different sizes will be oriented by a magnetic field. This is quantified through the order parameter along three orthogonal directions. The directions x, y and z are defined as follows: z is the direction perpendicular to the plane containing the molecular axes of the two distinct coronene molecules in a given unit cell. The y-axis is the direction bisecting the two. The angle between the two directors is 48.5° (β) or 87.4° (γ).

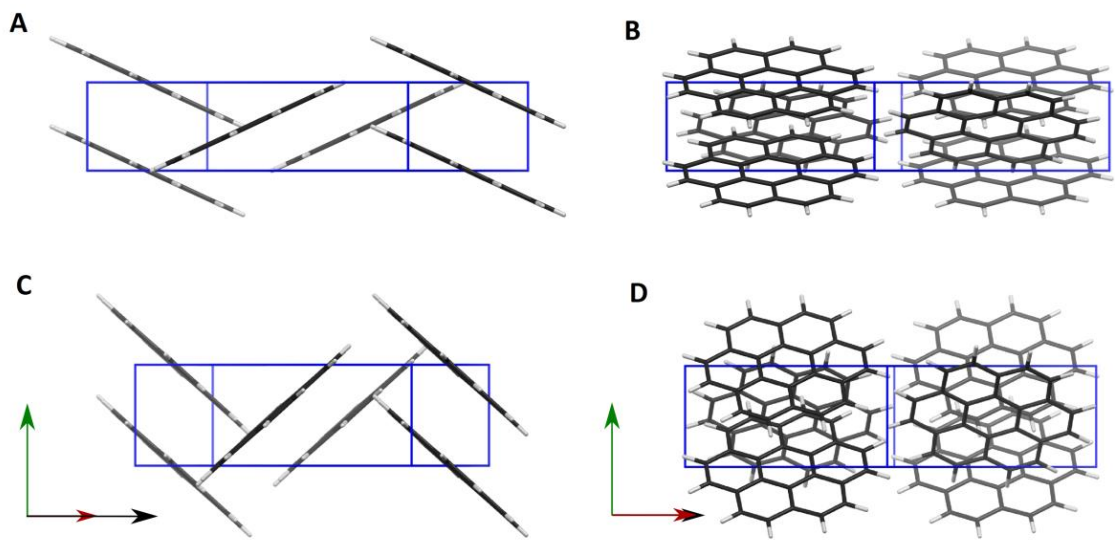

**Extended Data Figure 5. X-Ray diffraction crystal structures of β-coronene (A, B) and γ-coronene (C, D).** Red green and black arrows indicate the direction of the *a*-, *b*- and *c*-axis respectively.

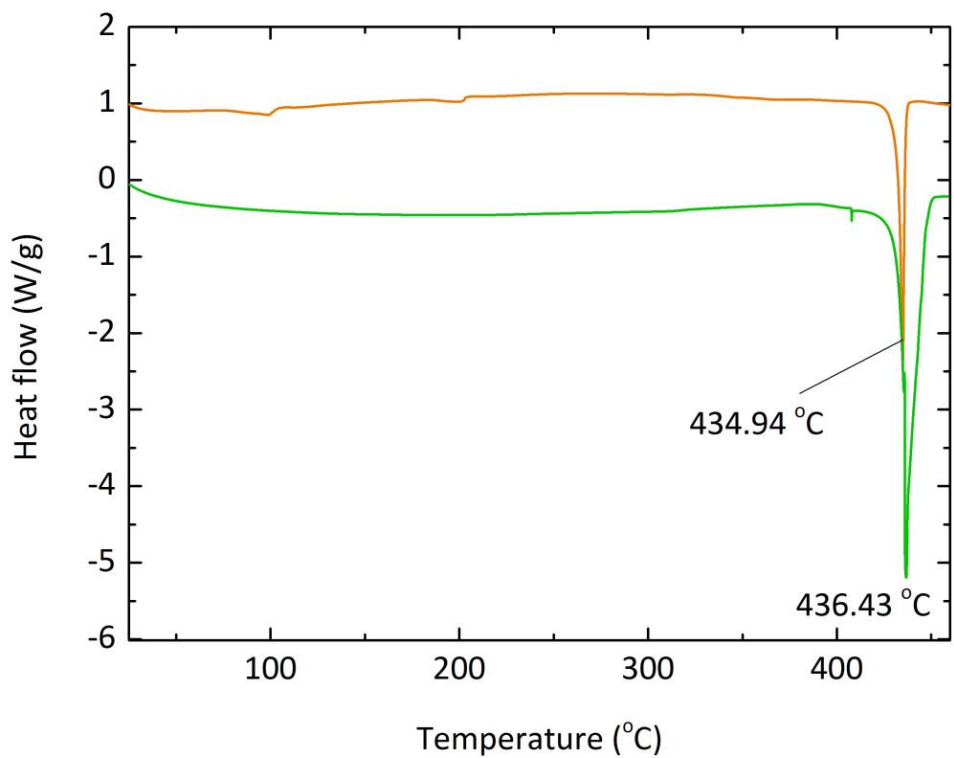

**Extended Data Figure 6. DSC curves** showing the melting profile of β-coronene (orange) and γ-coronene (green).

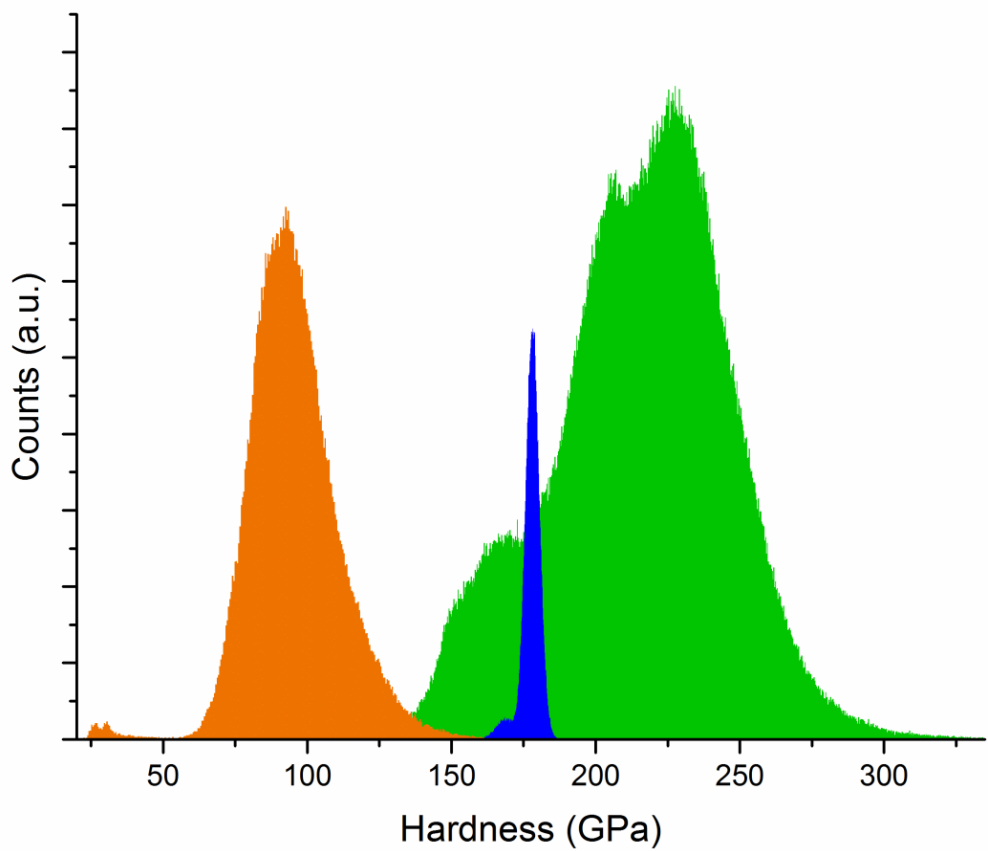

**Extended Data Figure 7. Elastic moduli measured via AFM.** The orange data are the results from β-coronene, the green from γ-coronene and the blue from a mica control sample. Coronene single crystals were measured on the $\overline{1}01\underline{1}$ and $10\overline{1}\overline{1}$ crystal faces.

| Code | vdW | form | $a$ [Å] | $b$ [Å] | $c$ [Å] | $\beta$ [°] | $d$ [Å] | ∠ [°] | $V$ [Å$^3$] | $E_\gamma$-$E_\beta$ [kJ/mol] |
|---|---|---|---|---|---|---|---|---|---|---|
| CASTEP | G06[a] | γ | 9.779 | 4.548 | 15.422 | 107.26 | 3.29 | 87.4 | 655 | 4.3 |
|  |  | β | 10.254 | 3.672 | 16.851 | 95.76 | 3.35 | 48.5 | 631 |  |
| CASTEP | GF[b] | γ | 9.924 | 4.665 | 15.915 | 105.53 | 3.52 | 82.1 | 710 | 3.8 |
|  |  | β | 10.336 | 3.871 | 17.264 | 95.04 | 3.58 | 44.9 | 688 |  |
| VASP | TS[c] | γ | 9.954 | 4.566 | 15.443 | 106.57 | 3.35 | 85.6 | 673 | 3.4 |
|  |  | β | 10.319 | 3.714 | 17.253 | 95.86 | 3.39 | 48.1 | 658 |  |

[a] Grimme (2006) – ref. (*50*).
[b] Grimme (2006) with experimental van der Waal radii as suggested by Fedorov *et al* (*51*).
[c] Tkatchenko-Scheffler – refs. (*53, 54*).

**Extended Data Figure 8.** Calculated structural parameters for coronene in the two crystal forms γ and β from DFT-D calculations using codes and dispersion corrections (vdW) as indicated. Lengths *a*, *b*, *c* are the lattice parameters and *β* the monoclinic angle; *d* is the interplanar distance between parallel coronene molecules. The columns labelled ∠ and V report the herringbone angle between adjacent coronene molecules and the unit cell volume, respectively. The final column reports the lattice energy difference between the two structures; positive indicates β is the more stable.

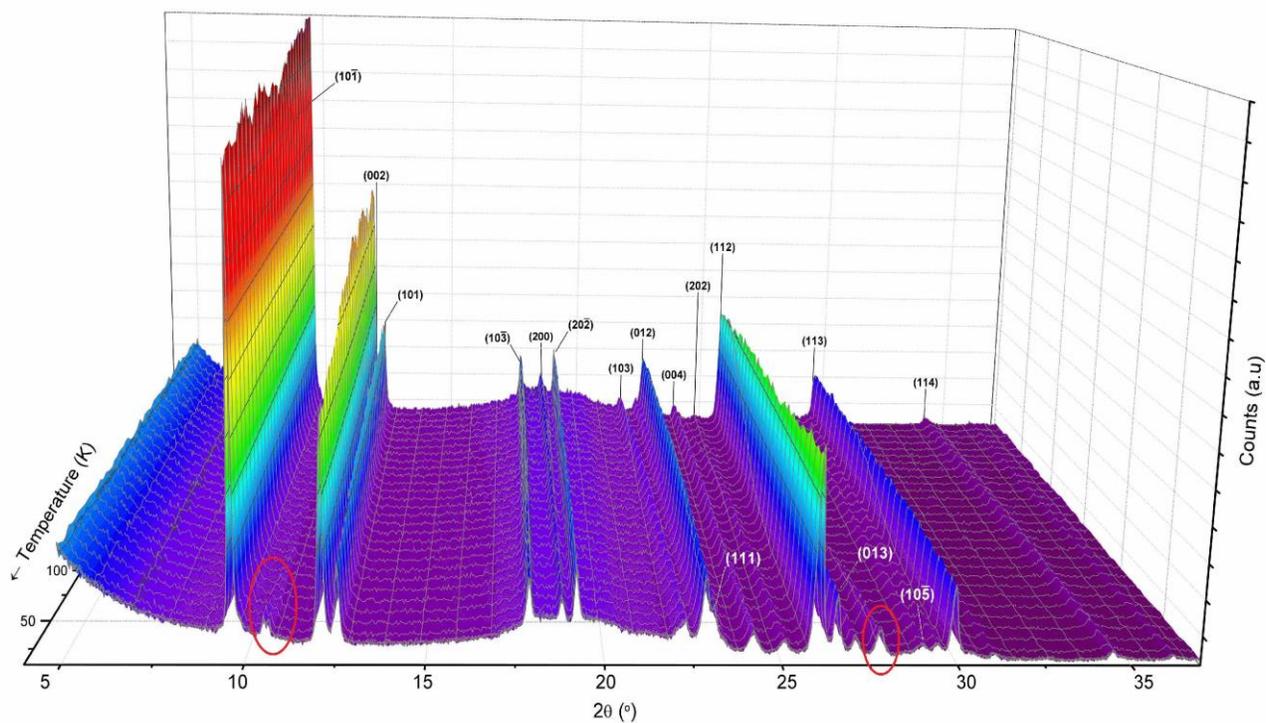

**Extended Data Figure 9. 3D surface plot of powder XRD as a function of temperature for γ-coronene** while cooling from 300 K to 12 K, without background correction. Arrow indicates forward direction of temperature. Color gradient is arbitrary used only for assisting with visualizing intensities. Red ovals identify emergent peaks due to the formation of β-coronene (viz. Fig. 6). The red oval near 10° 2$\theta$ encompasses two emergent peaks.

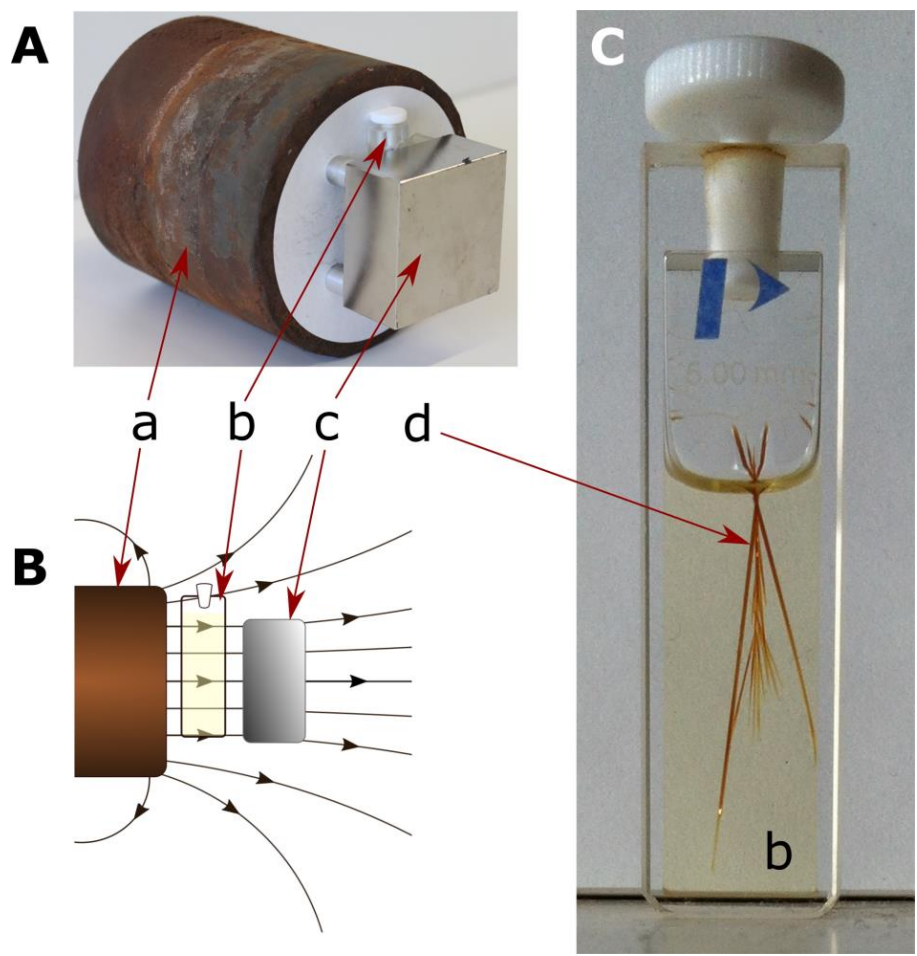

**Extended Data Figure 10. Experimental configuration for the growth of β-coronene.** (A) Optical image of the experiment, (B) schematic and (C) reaction vial with crystals of β-coronene *in-situ*. (a) is the 1 T magnet; (b) the sample vial; (c) the magnetic holding plate and (d) crystals of β-coronene. Field lines from the 1 T magnet through the experiment are indicated.

| Form | Temp. [K] | Code | *a* [Å] | *b* [Å] | *c* [Å] | $\beta$ [°] | d[Å] | ∠ [°] | V [Å$^3$] |
|---|---|---|---|---|---|---|---|---|---|
| γ | 150 | - | 10.02 | 4.67 | 15.06 | 106.7 | 3.43 | - | 699 |
| γ | 200 | - | 10.040 | 4.681 | 15.6041(19) | 106.32 | 3.43 | 85.76 | 703.77 |
| γ | 250 | - | 10.072(3) | 4.6907(12) | 15.650(6) | 106.18 | 3.44 | 85.59 | 710.1(4) |
| | | | | | | | | | |
| β | 80 | - | 10.386 | 3.821 | 17.211 | 96.24 | 3.47 | 49.7 | 679 |
| β | 150 | - | 10:392 | 3.839 | 17.229 | 96.24 | 3.48 | 50.0 | 683 |

**Extended Data Figure 11.** Experimental structural parameters for coronene in the two crystal forms γ and β. The parameters *a*, *b*, *c* are the three unit cell axis and *ß* the monoclinic angle, d represents the interplanar distance between parallel coronene molecules. The columns labelled ∠ and V report the herringbone angle between adjacent coronene molecules and the unit cell volume, respectively.